# Momentum selection for enhanced adaptive focusing through semi-transparent media.

Diego Di Battista[1,*], Giannis Zacharakis[1], Marco Leonetti[2]
1-Institute of Electronic Structure and Laser, Foundation for Research and Technology-Hellas, N. Plastira 100, Vasilika Vouton, 70013, Heraklion, Crete, Greece.
2-Center for Life Nano Science@Sapienza, Instituto Italiano di Tecnologia, Viale Regina Elena, 291 00161 Roma (RM) Italia.
∗To whom correspondence should be addressed; E-mail: dibattista.d@iesl.forth.gr.
November 2014

**Adaptive optics can focus light through opaque media by compensating the random phase delay acquired while crossing a scattering curtain. The technique is commonly exploited in many fields, including astrophysics, microscopy, biomedicine and biology. A turbid lens has the capability of producing foci with a resolution higher than conventional optics, however it has a fundamental limit: to obtain a sharp focus one has to introduce a strongly scattering medium in the optical path. Indeed a tight focusing needs strong scattering and, as a consequence, high resolution focusing is obtained only for weakly transmitting samples. Here we disclose an unprecedented phenomenon which allows to obtain highly concentrated optical spots even by introducing a minimum amount of scattering in the beam path that is with semi-transparent materials. By filtering the pseudo-ballistic components of the transmitted beam we are able to experimentally overcome the limits of the adaptive focus resolution, gathering light on a spot with a diameter which is one third of the original speckle grain.**

The active control of light propagation through turbid media[1] is becoming an essential tool in microscopy[2], biological and biomedical imaging[3-8], communication technology[9,10], and astrophysics[11,12]. Wavefront shaping[13] is a powerful technique that allows to manipulate the optical paths through scattering media and currently it is possible to generate behind a scattering material multiple light spots, actively driven at user controlled positions[1,14], spatiotemporal focusing[1,15], sub-wavelength foci (which may be employed for high resolution microscopy below the diffraction limit)[16,17], to transmit image around corners[18], to control nonlinear systems[6,19] such as random lasers[20-22] and to permit novel forms of secure communication[23].

These results are achieved by properly adjusting the wavefronts in order to correct for the de-phasing acquired due to the random propagation in the disordered medium. The key enabling technology is the Spatial Light Modulator (SLM), a device which enables a point per point control of the wavefront of a coherent light beam. In practice by setting with the SLM the input beam shape it is possible to control the wavefront at the output of an optical system with an unknown scattering matrix. Various strategies have been developed to obtain light focused at a user defined location: time reversal phase conjugation[24,25], transmission matrix measurement[26,27] or phase scan based algorithms[28-30]. Vellekoop and colleagues demonstrated that the minimum spot size achievable through adaptive focusing in strongly scattering materials is defined by the speckle correlation function[14]: in practice its limit is the speckle grain. The speckle pattern is a random distribution of bright and dark areas due to random interference of countless light paths transmitted through disorder and its grain size depends on the length $L$ and transport mean free path $\ell$[31-33]: in transmission geometry, when the scattering strength increases, the typical grain of the speckle pattern becomes smaller. In order to obtain a tight focus through a scattering sample one has to exploit thick samples because high resolution is obtained in exchange of throughput[34], which is a critical obstacle for modern adaptive super-resolution techniques[16,17], currently limited to low transmittance experiments only. Optical Eigenmode (OEi) approaches were tested to achieve sub-diffraction optical features in free space for minimizing the size of a focused optical field[35]. The combination of particular photonic structures and wavefront correction by OEi methods have been exploited to produce subwavelength foci[36,37]. However, to date and to the best of our knowledge, adaptive foci with a resolution under the limit of the speckle grain size have never been demonstrated.

The current optical techniques adopted for bio-imaging are efficient enough up to the first millimeter in depth (1 transport-mean-free-path)[4] and wavefront modulation appears to be the best solution for correcting imaging quality[3,38]. Here we propose an experimental technique which permits to achieve sharp adaptive focusing through weakly scattering samples. A single speckle pattern results from the superimposed contribution of many different components. We demonstrate that by appropriately selecting some of the components during the optimization we can produce a focus smaller than the speckle grain, effectively overcoming the theoretical limit previously proposed by Vellekop and coworkers[14] for strongly scattering media. The core idea consists of selecting only those light paths which experienced multiple scattering events by filtering the ones which have experienced only weak scattering. A spatial filter is employed for the selection of the appropriate light paths (modes)[39] and by exploiting a standard phase scan method[30] (see Methods) a stronger focusing with a smaller size speckle pattern is obtained. Furthermore the focus persists also if the filter is removed so that a sharp light spot is obtained within a speckle pattern with a much larger grain. By exploiting our protocol we are able to obtain a focus size approximately 68% smaller than the average speckle grain.

## Results

**Adaptive focusing with Spatial High Pass filter.** The mode filtering relies on custom-made Spatial High Pass filters (HP filter) with diameters ranging from 0.35mm up to 1.3mm (see Methods). In Fig. 1 a schematic representation of the experimental setup is shown: a SLM modifies the wavefront of the coherent light which impinges onto a scattering sample (S). During propagation through the turbid material, scattering decomposes an incident wave into multiple components which generate a speckle pattern by randomly interfering at the output of the sample. The speckle pattern is collected by an objective lens (OBJ) (see Methods for further details) and is projected behind it. At a distance of 150mm from the rear face of the OBJ a lens L5 is placed to reproduce the speckle pattern (and a real image of the sample) at its focal length (see Fig. 1). On this plane the HP filter is aligned in order to block the central components of the speckle pattern: these components are related to modes which underwent a few scattering events (hence they are weakly scattered from the ballistic trajectory; we refer to them as *"pseudo-ballistic"*). Finally a 400mm lens L6 produces an image on the CCD camera plane which is the conjugate plane of the image from L5. The image on the camera is the result of the superposition between modes not blocked by the HP filter or by the Polarizers (P1 and P2 with perpendicular orientations serve to eliminate ballistic contribution). Different samples were illuminated with the same beam waist (0.24mm) and the back surface of each sample has been aligned on the focal plane of the OBJ. We define the Full Width at Half Maximum, FWHM (*w*) as the width of the intensity peak around the target spot.

The optimization protocol starts from a random SLM mask, splitting the wavefront into segments with random de-phasing. Then for each one of the segments the phase is shifted (see Methods) and the change is retained if the brightness at the target is enhanced, otherwise the previous configuration is restored. Our experimental setup allows for an *enhancement* factor ($\eta$)[13,14] of 250, calculated as the ratio between the intensity at the target and the average intensity of the speckle pattern before optimization. In Fig. 2 we compare the values *w* obtained with the standard focusing approach[13,14] (without filtering) for different samples characterized by the optical length $L/\ell$ (blue circles in Fig. 2) with the values obtained when the HP filter is used (red squares in Fig. 2). We study the focusing resolution, measured as the full width at half maximum *w* for different scattering samples with controlled optical transmittance (sample thickness is varied from a few micrometers to hundreds of micrometers) corresponding to optical lengths between one and ten transport mean free paths[33] (from $L/\ell$=0.7 to $L/\ell$=6.5). The ratio $L/\ell$ was calculated from the measured direct transmission exploiting a modified Beer-Lambert law[33,40] (see Methods). The results of Fig. 2 obtained for samples with different optical thicknesses $L/\ell$, and presented by the blue circles, demonstrate that turbid lenses reach the optimum efficiency for an optical length of the order of the transport mean free path.

The results of the same experiment with a HP filter applied as presented in Fig. 2 by the red squares, demonstrate that the FWHM of the foci obtained with the filter blocking the pseudo-ballistic modes (see sketch Fig. 2a) is always smaller than the non-filtered case.

**A Sharp Adaptive focus in a semi-transparent medium.** We characterized the effects of the HP filters demonstrating strong adaptive focusing through semi-transparent materials. We measured the focus width, *w,* as a function of the HP filter diameter *D,* through a semi-transparent sample.

Fig. 3 reports the results for *D* varying from 0.35 to 1.3 mm. *w* decreases by increasing the filter diameter and it is possible to obtain a focus much smaller than the size of the speckle grains. Panel a) on Fig. 3 shows the speckle pattern at the back of the disordered sample (optical length $L/\ell$=0.81). Panels b) and c) show the same pattern filtered by a 0.7mm and a 1.3mm diameter beam stop, respectively. By eliminating the pseudo-ballistic components, our filter selects modes which have suffered stronger scattering and thus produce a larger effective numerical aperture yielding a smaller speckle and a smaller focus. On average (statistics over 10 measurements), we obtained a focus which is of 0.32±0.15 (one third) of that obtained when the filter is absent.

Fig. 4a presents a snap-shot of the initial speckle pattern generated through a semi-transparent sample, before the optimization, when the HP filter is not inserted; the pattern composed of large grains with an average diameter of 25µm estimated from the speckle correlation function. In Fig. 4b we report an image of the focus achieved with a 0.5mm HP filter. A remarkable effect is observed after the removal of the HP filter; the adaptive focus is not affected and remains smaller than the speckle size as shown in Fig. 4c.

The filtering of the pseudo-ballistic modes is only needed during the optimization procedure, while the sharp focusing is retained after the filter removal. In Fig. 4d we compare the foci profiles. The solid curve correspond to the measurement obtained with the HP spatial filter, while the blue dashed curve corresponds to the standard focusing approach[13,14] without filtering. In this case the effect of the filter is a reduction of the focus size by a factor of 0.52±0.11; moreover, removing the spatial filter does not alter the focus, which maintains the same shape, while being surrounded by larger speckle grains. The focus waist obtained is 12.3±0.8μm (blue profile in Fig. 4e) which is approximately the same as the size of the focus obtained with the filter (FWHM is of 12.1±0.8μm, represented by the blue dashed curve in Fig. 4d and 4e) and is approximately half of the speckle size grains (FWHM of the speckle correlation function is 23.5±2.5μm, represented by the red curve in Fig. 4e). When the filter is absent the presence of the pseudo-ballistic modes increases the background signal with respect to the focus intensity decreasing the Peak-to-Background Ratio *($\eta_{PBR}$)* calculated as the maximum intensity at the target divided by the average intensity of the background. We measured *$\eta_{PBR}$* equal to 196±35 with filter and equal to 16±5 without filter (statistics over 10 measurements). The pseudo ballistic modes (with small numerical aperture) hide the modes associated with strongly scattering light paths (which produce a less intense contribution but a larger effective numerical aperture) making impossible to obtain a sharp focus when both the contributions are interfering on the image plane.

**Discussion**

We have demonstrated the enhanced adaptive focusing through weakly scattering media, by the introduction of a spatial filter in the image plane of the produced speckle pattern. The effect is a significant increase of the focusing resolution with a reduction of the spot size below the speckle size defined by the speckle correlation function. The method strongly improves the focusing resolution for turbid samples (turbid lenses) with optical lengths smaller than two transport mean free path. The introduction of a spatial filter reduces the speckle size because pseudo-ballistic modes possess a reduced span of momentum components (thus producing a larger effective numerical aperture) with respect to strongly scattered modes. This means that in the weak scattering regime, speckles result from a superposition of patterns possessing different grain size. Our filtering technique, selects a speckle with a smaller size during the optimisation protocol, and this allows the algorithm to exploit degrees of freedom which are "hidden" in the standard focusing optimisation. In fact, in the extremely weak scattering configuration, pseudo-ballistic modes are much more intense than the strongly scattering modes, thus producing a large grained speckle which hides the small grained speckles produced by the strongly scattering channels. With our approach we are thus able to select the components of a speckle pattern producing a focus with the maximum resolution. The presented results not only implement the state-of-art of the adaptive focusing process but may also open the way to a novel generation of high transmission, semi-transparent turbid lenses with high focusing resolution.

**Acknowledgements**

This work was supported by the Grants "Skin-DOCTor" and "Neureka!" implemented under the "ARISTEIA" and "Supporting Postdoctoral Researchers" Actions respectively, of the "OPERATIONAL PROGRAMME EDUCATION AND LIFELONG LEARNING", co-funded by the European Social Fund (ESF) and National Resources and from the EU Marie Curie Initial Training Network "OILTEBIA" PITN-GA--2012-317526. The authors thank Stella Avtzi for helping in HP filters fabrication and characterization.

**Author Contributions**

All authors have contributed to the development and/or implementation of the concept. D.D.B. and M.L. ideated and designed the experiment. D.D.B. realized the experimental apparatus and performed the measurements. G.Z. analyzed the data and supported the development of the experiment.

**Additional information**

Competing financial interests: The authors declare no competing financial interests.

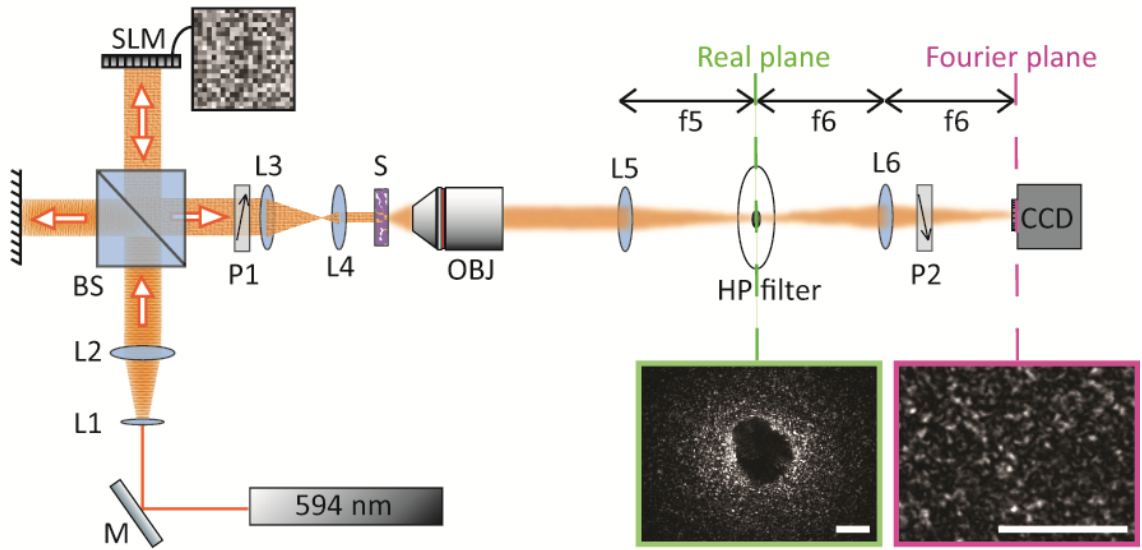

*Figure 1 A schematic representation of the experimental setup: a modulated beam from a spatial light modulator is reduced and collimated by a 15X telescope and it impinges onto a scattering layer (S). The transmitted light is used to produce a real image of the sample by lens L5. The image is filtered from its central components by a Spatial High Pass filter (HP filter). The result of the filtering is Fourier transformed onto the camera plane by the lens L6. Polarizers P1 and P2 have perpendicular orientation in order to filter the ballistic contribution. Scale bars correspond to 0.5mm.*

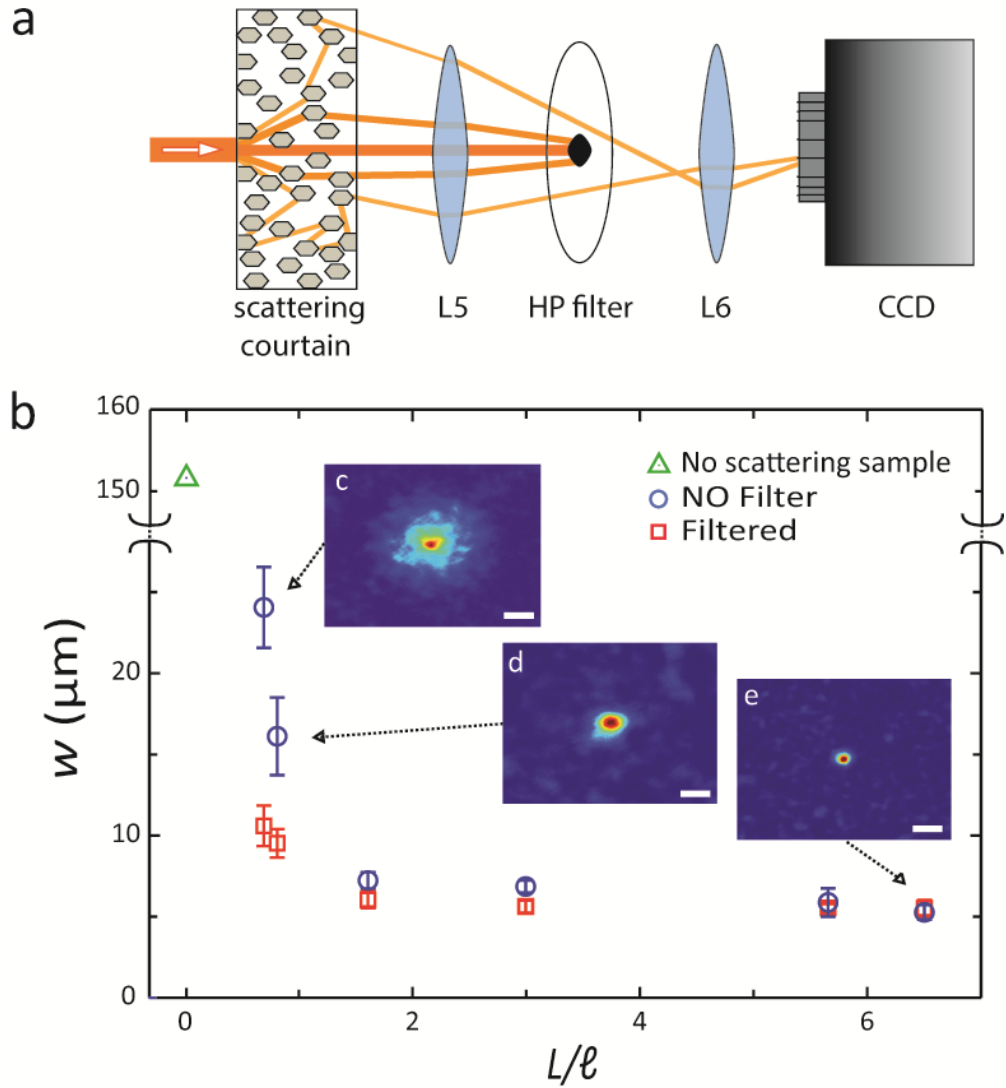

*Figure 2:* a) Graphical representation of the operating principle of the HP filter: light rays propagating through a weakly scattering medium are focused on the plane of the HP filter. Only light that experiences few scattering events is blocked by the filter, the remaining contributes to the speckle at the camera plane. b) Full Width at Half Maximum of foci obtained at the end of the optimization process plotted as a function of the optical length $L/\ell$ of the samples. The green triangle represents the beam waist that impinges onto the sample. A comparison between the ordinary approach[14] (no filter, blue circles) and the case with the 0.8mm diameter beam stop filter (red squares) is shown. The three insets, starting from the top, correspond to foci obtained at $L/\ell$=0.7, 0.81 and 6.3. White scale bars correspond to 30µm.

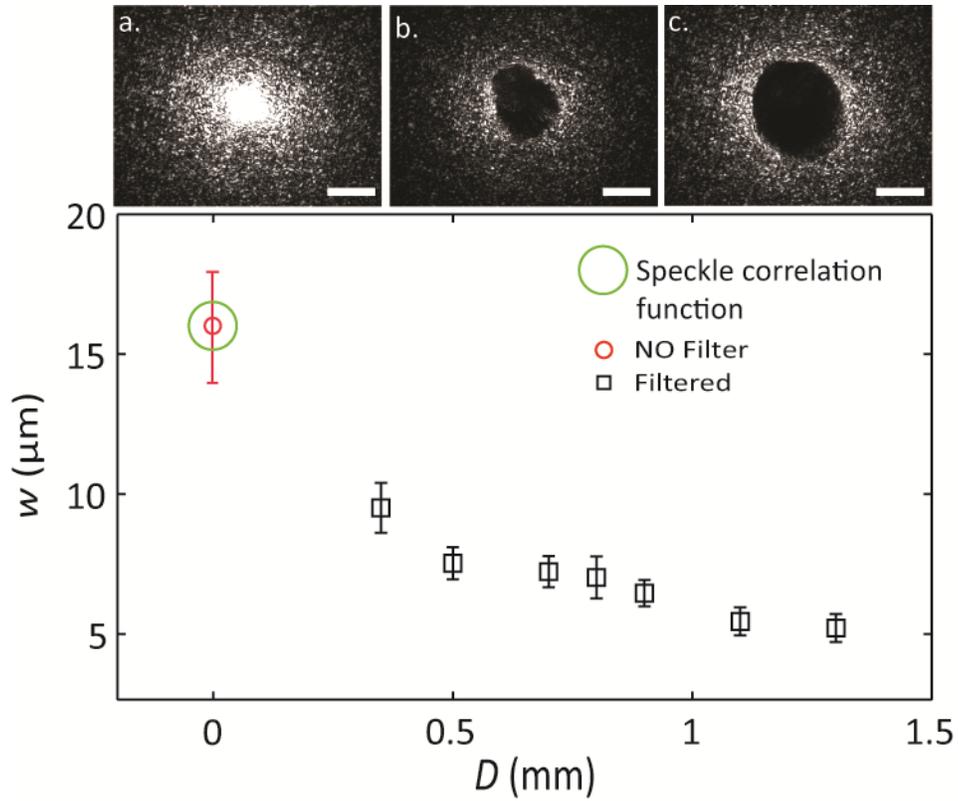

*Figure 3:* In the upper panel we report the speckle pattern in the plane generated by L5 (see Fig. 1) under three different configurations. Panel a. is the free speckle pattern through a sample with optical length $L/\ell=0.81$; b. and c. are filtered patterns with 0.7mm and 1.3mm diameter spatial filter, respectively. Scale bars are equivalent to 0.5mm. The graph shows the dependence of the focus width w (in black square) versus the beam stop filter diameter. All data has been collected through a sample with $L/\ell=0.81$ and is compared to the FWHM of the focus obtained in the absence of a filter (red circle).

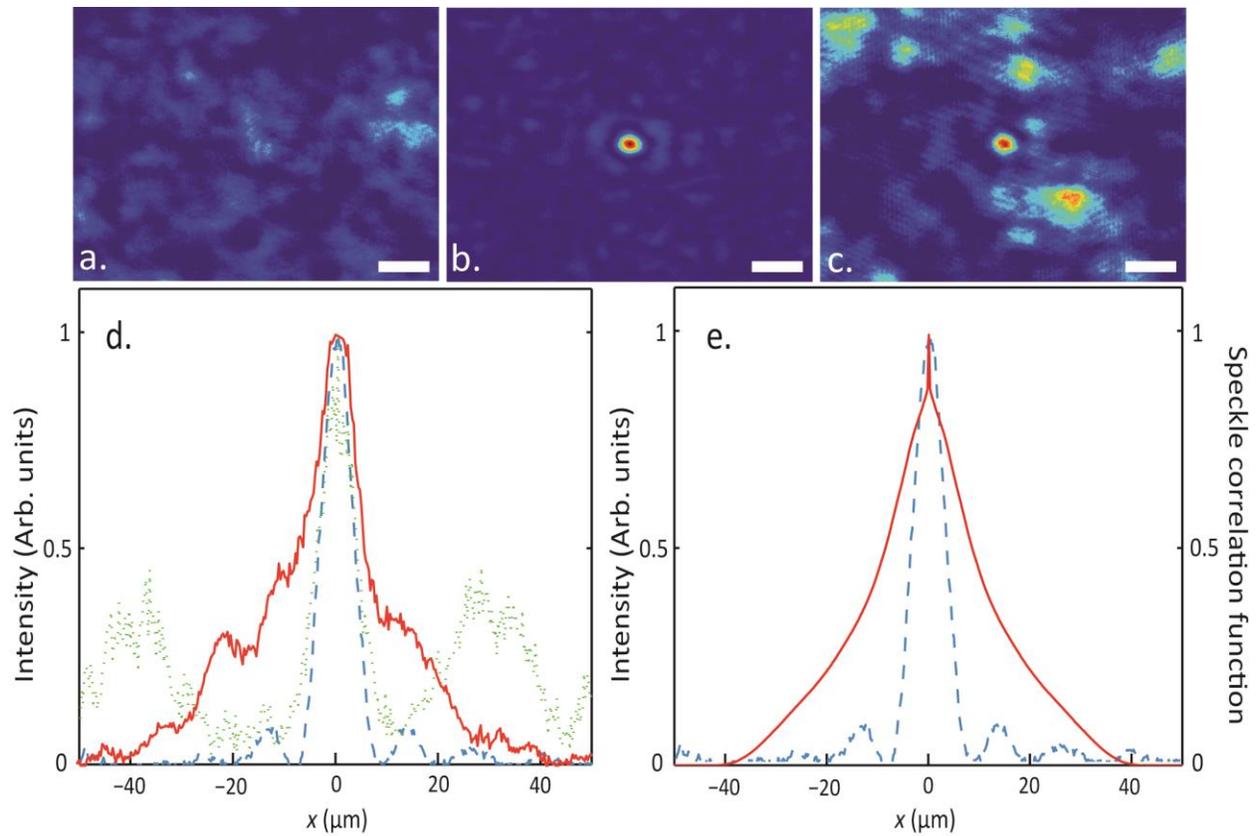

*Figure 4:* a) speckle pattern obtained through a $L/\ell=0.7$ scattering sample. b) shows the optimized focus with a 0.5mm diameter filter for the same sample. c) shows the same as in b but after the filter is removed. Note: the focus still present and smaller than the speckle grain. White scale bars correspond to 30μm. d) Focusing profiles at the target position without filter (solid red curve), with filter (dashed blue curve) and when the filter is removed (dotted green curve). e) The focus obtained with the filter (dashed blue curve) is compared to the speckle correlation function with a randomly generated wavefront and in the absence of a filter (solid red curve).